# Free-Space Imaging Beyond the Diffraction Limit Using a Veselago-Pendry Transmission-Line Superlens

Ashwin K. Iyer, *Graduate Student Member*, *IEEE* and George V. Eleftheriades, *Fellow*, *IEEE*

*Abstract*— Focusing using conventional lenses relies on the collection and interference of propagating waves, but discounts the evanescent waves that decay rapidly from the source. Since these evanescent waves contain the finest details of the source, the image suffers a loss of resolution and is referred to as 'diffraction-limited'. Superlensing is the ability to create an image with fine features beyond the diffraction limit, and can be achieved with a 'Veselago-Pendry' lens made from a metamaterial. Such a Veselago-Pendry superlens for imaging in free space must be stringently designed to restore both propagating and evanescent waves, but meeting these design conditions (isotropic $n=\varepsilon_r=\mu_r=-1$) has proven difficult and has made its realization elusive. We demonstrate free-space imaging with a resolution over three times better than the diffraction limit at microwave frequencies using a Veselago-Pendry metamaterial superlens based on the negative-refractive-index transmission-line (NRI-TL) approach, which affords precise control over its electromagnetic properties and is also less susceptible to losses than other approaches. A microwave superlens can be particularly useful for illumination and discrimination of closely spaced buried objects over practical distances by way of back-scattering, e.g. in tumour or landmine detection, or for targeted irradiation/hyperthermia.

*Index Terms*—metamaterial, left-handed (LH), negative refractive index (NRI), focusing, transmission line, periodic structures, superlenses, super-resolution, diffraction limit.

## I. INTRODUCTION

The long-held interest in synthesizing known material properties artificially has been revived by the impressive recent developments in metamaterials, which are artificial materials engineered to exhibit electromagnetic phenomena not available or not readily available in nature, such as a negative permeability ($\mu$) and a negative permittivity ($\varepsilon$), as well as their most unusual product: a negative refractive index (NRI). Through the theoretical work of V. Veselago [1] and J. B. Pendry [2], it is known that flat NRI lenses designed to meet certain strict conditions are able to focus the propagating-wave components of a source without geometric aberration while simultaneously restoring the amplitude of its evanescent-wave components, which decay quickly as they depart the source, such that the focal plane contains an exact real image of the source, down to its finest features. As a result, such a lens, appropriately termed a 'Veselago-Pendry superlens,' is able to overcome the constraints of the classical diffraction limit, which restricts focusing with conventional lenses to a resolution on the order of half the wavelength of illumination. Practical Veselago-Pendry superlenses have many potential biomedical, microelectronics, and defense-related applications in sub-diffraction microscopy, lithography, tomography, and sensing.

Although such a lens has been realized using negative-refractive-index transmission-line (NRI-TL) metamaterials for embedded sources in planar and three-dimensional form [3, 4] and inside waveguide environments [5, 6], it has, so far, eluded practical realization in a form capable of interacting with sources in free space, the form in which it was first envisioned. This is largely due to the fact that a true free-space Veselago-Pendry superlens has a number of stringent design requirements: first, the lens must possess $\mu = -\mu_0$ and $\varepsilon = -\varepsilon_0$ for the polarization(s) concerned (where $\mu_0$ and $\varepsilon_0$ are the free-space permeability and permittivity, respectively) in order to be impedance-matched to free space and simultaneously possess an effective refractive index $n = -1$, which also renders it aberration-free. Since these materials are necessarily dispersive, achieving these values with adequate precision requires a means of tightly controlling the metamaterial's frequency response. Second, the lens must be extremely low loss and be adequately thin, since both loss and electrical thickness serve to quickly degrade the resonant evanescent enhancement contributing to sub-diffraction imaging. A third condition follows from the previous: the unit cells comprising the lens must themselves be deeply sub-wavelength in size in order to minimize spatial anisotropy and to ensure that the structure possesses the desired bulk response. Last, the transverse dimensions of the lens must be large enough that the lens can be illuminated by a source in free space.

True Veselago-Pendry superlenses (that is, metamaterials capable of restoring both the propagating and evanescent spectrum of a source) have not yet been realized for sub-diffraction imaging in free space, but many other varieties of metamaterial have experimentally demonstrated phenomena akin to sub-diffraction imaging by associated physical mechanisms. These include the plasmonic silver film [7], the magneto-inductive lens [8], and the swiss-roll structure [9] which, although they do not possess a negative refractive index and so cannot focus the propagating wave numbers, do recover fine spatial features through evanescent enhancement; unfortunately, this means that such lenses impose an extremely short working distance (typically less than $\lambda_0/8$, where $\lambda_0$ is the free-space wavelength) between the source, lens, and image. Sub-diffraction imaging phenomena have been successfully extended to the far field using magnifying 'hyperlenses' [10, 11, 12], but this class of superlenses relies explicitly on anisotropy and sources are typically applied directly to the hyperlens face; as a result, the working distance on the source side remains limited. Sub-diffraction imaging using printed split-ring-resonator- (SRR-) based structures has been reported in the way of transversely and longitudinally confined sub-wavelength focal 'spots' [13, 14], in spite of their high losses and/or large electrical thickness; however, calculations based on [15] suggest that the reported loss, lens thickness, and observed resolution ability are inconsistent, if the structures are to be regarded as true Veselago-Pendry superlenses. Furthermore, the lenses described in these works employ a refractive index of −1.8, which defies even the basic requirement that the refractive index of a free-space Veselago-Pendry superlens be −1 in order to restore the propagating spectrum of the source and cause the resonant enhancement of the evanescent-wave components to degenerate to the same frequency; contrary to the authors' claim in [13], the resolution ability of a Veselago-Pendry superlens is not tolerant to such large deviations in its effective parameters. Moreover, the use of a higher index reduces the working distance between lens and image. Last, the authors' use of large (resonant) antennas to detect the fields precludes the observation of the true evanescent field magnitudes, which necessarily manifest themselves only in a transverse (and not longitudinal) confinement of the fields at the focal plane [3] – in fact, as discussed in [16], the use of such large antennas strongly perturbs the fields, making the observed resolution more a function of the antenna measurement than of the fields produced by the metamaterial. Thus, the results described in these works are inconsistent with the imaging principles of the Veselago-Pendry superlens and the requirements outlined above; indeed, in attempting to explain these inconsistencies, the authors of [13, 14] have speculated that anisotropy, rather than sub-diffraction imaging by way of the restoration of evanescent waves, may be responsible for these phenomena [14]. Although these factors preclude their description as Veselago-Pendry superlenses, further research into such structures may reveal other intriguing mechanisms by which sub-diffraction imaging can be achieved.

The first successful attempt at superlensing [3], although in a planar transmission-line (TL) environment rather than a free-space environment, employed NRI-TL metamaterials, which consist of a fine TL grid loaded with inductors and capacitors to control their effective-medium response [17]. Subsequently, it was considered that a 'volumetric' NRI-TL metamaterial could be realized for free-space excitation by stacking planar NRI-TL metamaterials in a multilayer fashion [18, 19]. While not three-dimensionally isotropic, such a structure would appear isotropic to two-dimensional excitations (i.e. an infinite line source) in free space and could be fabricated easily and rapidly using widely available printed-circuit-board (PCB) technologies. Previously, such a structure was realized using fully printed loading elements (interdigitated capacitors and meandered inductors, and without any vias) and demonstrated diffraction-limited focusing of a free-space magnetic dipole source consistent with the use of large unit cells and an electrically thick lens [20]. However, it was also suggested in that work that the unit cells may be made simultaneously low-loss and electrically small by exploiting the strong lumped loading afforded by discrete chip inductors and capacitors with high quality factors, as in the planar case, which would also enable the realization of an adequately thin, NRI-TL metamaterial, free-space superlens. This NRI-TL superlens, and the resulting experimental verification of the phenomena of sub-diffraction focusing and super-resolution in free-space, is the subject of this work. These results show that the so-far elusive free-space Veselago-Pendry superlens is, indeed, realizable, and arguably represent the first such realization.

## II. DESIGN

The NRI-TL approach is known to offer intrinsically large NRI bandwidths and minimize losses through the tight coupling between unit cells [21]. The strength of the lumped elements loading the host TL medium renders the unit-cell size deeply sub-wavelength, allows the realization of an adequately thin Veselago-Pendry lens, affords precise control over its effective-medium properties, and mitigates losses. The TL host medium constituting the layers of the volumetric NRI-TL lens employs a topology known as the 'series' NRI-TL node, which intersects four 1D NRI-TLs in series and so obviates the need for vias [18]. The most suitable TL for this topology is the fully uniplanar co-planar strip (CPS) TL. When appropriately loaded by inductors and capacitors in the NRI-TL configuration [17], this topology may alternately be viewed as a uniplanar array of capacitively loaded rings connected to each other using inductors. The substrate medium was chosen to be 0.5-oz.-copper-clad Rogers RO3003 60-mil (1.524mm) microwave substrate with $\varepsilon=3\varepsilon_0$ and $\tan\delta=0.0013$. The surface-mount elements selected were Coilcraft high-Q ultrasmall wirewound inductors (19.3nH at 2.4GHz) and American Technical Ceramics high-Q capacitors (1.2pF). The particular geometrical features of the ring were decided to accommodate the mounting patterns of the surface-mount components suggested by their manufacturers, and are shown in Fig. 1.

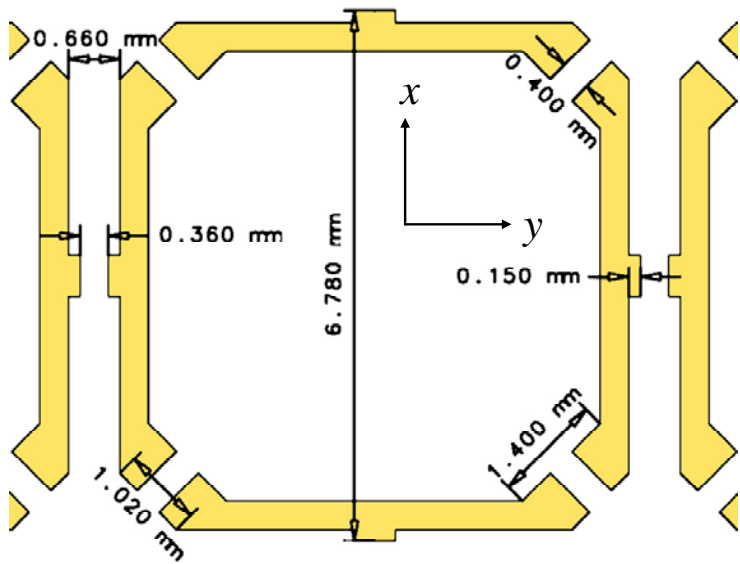

Figure 1: Relevant dimensions of co-planar strip (CPS) transmission-line (TL) host medium.

The optimal vertical layer spacing of 3.476mm was decided in concert with an equivalent-circuit model [18, 19] that accurately describes the effective-medium properties of the metamaterial to yield effective-medium parameters $\mu=-\mu_0$ and $\varepsilon=-\varepsilon_0$ at an operating frequency of 2.40GHz ($\lambda_0$ = 125mm in free space) when illuminated by *p*-polarized fields propagating in the layer planes (i.e., *z*-polarized magnetic fields). The unit cells in the present design were chosen to be nearly $\lambda_0/18$ in size (7.14-mm square), and so a lumped description is justified.

## III. SIMULATION

Simulations of the designed volumetric NRI-TL topology were performed using Ansoft's HFSS (High-Frequency Structure Simulator) [22], which is a full-wave finite-element-method electromagnetic field simulator. The ultimate goal of extracting the effective-medium parameters of a periodic medium requires knowledge of its intrinsic effective propagation constant, $\beta$, and wave impedance, $\eta$. The former may be obtained as a function of frequency by examining the spectrum of resonances within a single unit cell that are associated with particular periodic boundary-phase conditions. In HFSS, this type of simulation is known as an 'eigenmode simulation,' which can yield $\beta$ in a particular direction of propagation as a function of $\omega$ (the 'dispersion diagram'), or $\beta$ over all directions of propagation at a particular frequency (the 'iso-frequency' or 'equi-frequency' surface). The wave impedance may be obtained in a TL environment of known geometry (e.g. parallel-plate waveguide) by way of its characteristic impedance, $Z_0$. In HFSS, this, along with a measure of $\beta$, is obtained through the scattering (S-) parameters produced by a slab of the periodic medium that is finite (i.e., consisting of a finite number of unit cells) in the direction of propagation, but may be rendered infinite in other directions through the use of appropriate periodic boundary conditions. This is known as a 'driven simulation,' since a practical source (e.g., a normally incident plane wave) is used. By associating the obtained S-parameters with a homogeneous slab of equal thickness, its effective homogeneous properties can be obtained by way of an extraction procedure that is valid for electrically thin slabs [23]; certainly, the volumetric NRI-TL lens falls into this category.

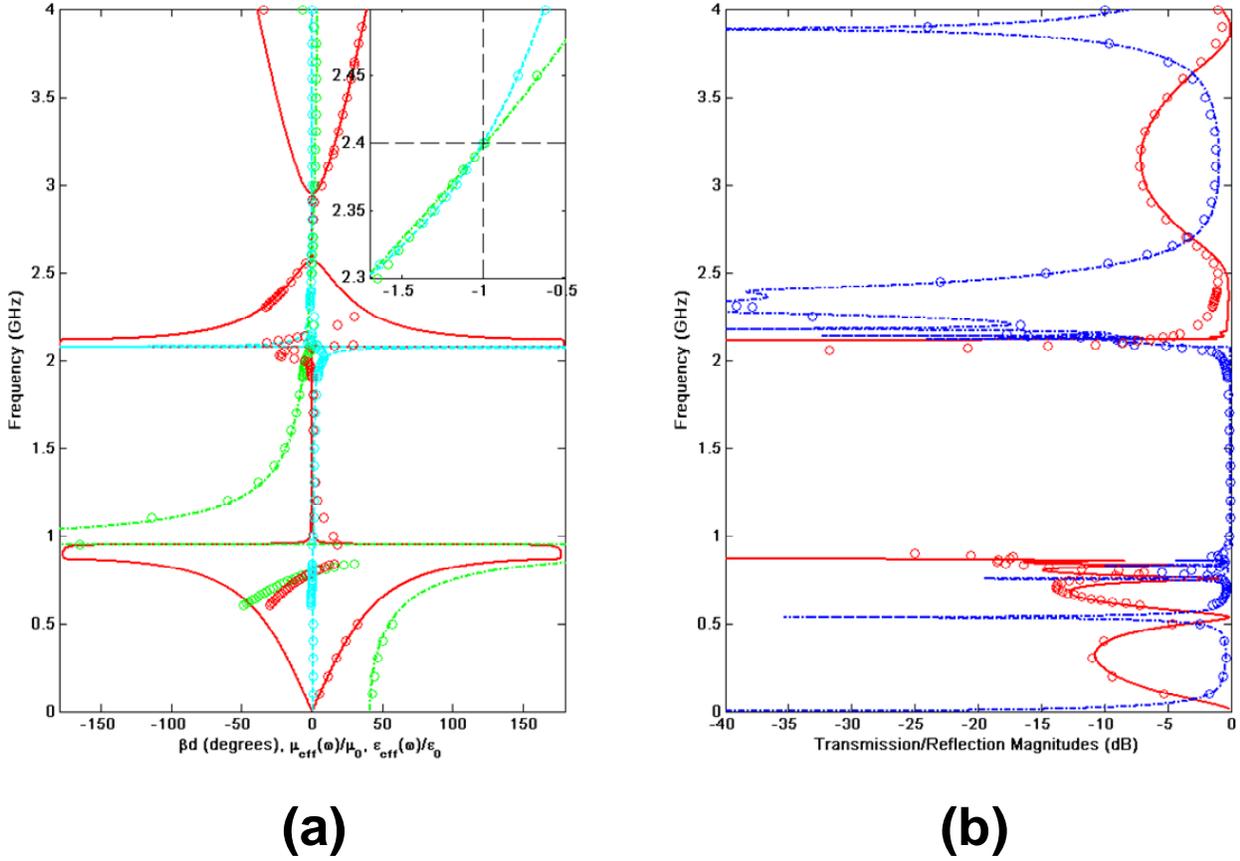

Figure 2: Effective material parameters and transmission/reflection magnitudes of a five-unit-cell-thick volumetric NRI-TL metamaterial, extracted from full-wave FEM simulation data (full-wave simulations – circles, equivalent-circuit model – curves): A) $\mu_{eff}(\omega)/\mu_0$ (dashed cyan curve and circles); $\varepsilon_{eff}(\omega)/\varepsilon_0$ (dash-dotted green curve and circles); $\beta d$ (solid red curve and circles). The inset shows the superlensing condition $\mu_{eff}(\omega)/\mu_0 = \varepsilon_{eff}(\omega)/\varepsilon_0 = -1$ near 2.4GHz; B) transmission magnitude $|S_{21}|$ (solid red curve and circles); reflection magnitude $|S_{11}|$ (dashed blue curve and circles).

In all cases, the lumped components were modelled by current sheets endowed with the appropriate lumped-element boundary conditions; dielectric loss tangents and component losses (as specified in their data sheets in terms of either equivalent series resistance or quality factor) were included, and the metallic features were specified as copper with a bulk conductivity reduced by over 70% from the nominal value to account for surface roughness, by way of HFSS's 'finite-conductivity' boundary condition.

The band structure, transmission magnitude, and transmission phase for propagation of a normally incident plane wave through a five-unit-cell-thick volumetric NRI-TL metamaterial slab are shown in Fig. 2. Also shown are the relative effective permittivity and permeability extracted from the S-parameters. These are represented by circles superposed on the theoretical curves obtained from the volumetric NRI-TL equivalent-circuit model described in [18, 19]. These data prove the excellent agreement between the two, especially where the phase shift per unit cell $\beta d$ is small (corresponding to the effective-medium limit in which effective material parameters $\mu_{\text{eff}}(\omega)$ and $\varepsilon_{\text{eff}}(\omega)$ are defined). It should be noted that there are certain frequency regions in which the extracted $\mu_{\text{eff}}(\omega)$, $\varepsilon_{\text{eff}}(\omega)$, and $\beta(\omega)$ seem to differ from the equivalent-circuit theory; these are not discrepancies – rather, they are frequency regions in which the extraction methods fail due to the resonances that cause large phase shifts per unit cell. The NRI region, exhibiting a prominent backward-wave characteristic (opposite phase and group velocities), is evident between approximately 2.1GHz and 2.6GHz, representing a fractional NRI bandwidth of over 21%. It is also apparent from the reflection magnitude ($S_{11}$ – dashed blue curve and circles) that the metamaterial is very well matched over this frequency range, and particularly well matched near 2.4GHz, where the return loss is better than –35dB. Although not depicted in the figure, the NRI metamaterial dispersion curve intersects the (PRI) light line at 2.39GHz (a difference of 0.4% compared to the design frequency of $f_0$=2.4GHz, but well within tolerance due to discretization of the FEM mesh) which, along with good matching and the electrical thinness of the slab at this frequency, uniquely suggests that the metamaterial possesses effective permittivity and permeability values $\mu_{\text{eff}}(\omega)=-\mu_0$ and $\varepsilon_{\text{eff}}(\omega)=-\varepsilon_0$ at this frequency. Indeed, the extracted permeability and permittivity, presented in the inset around 2.4GHz, correspond to the equivalent-circuit theory and intersect at relative values of –1 at 2.39GHz. The insertion losses in the passband, although larger in simulation than in the theoretical analysis, remain near 0.2dB per unit cell, which is consistent with the use of low-loss materials and lumped components with high quality factors, and also with the requirements of the Veselago-Pendry superlens. This should be compared to an insertion loss of over 2dB per unit cell for other metamaterials reporting similar characteristics (see, for example, Fig. 1 in [14]).

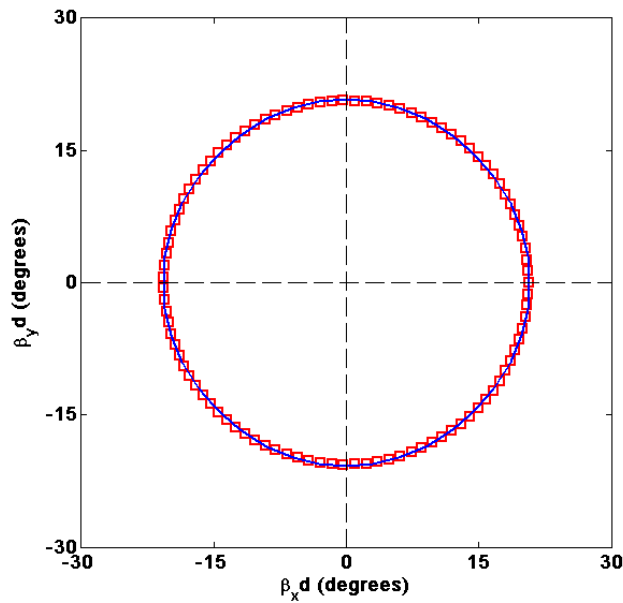

Figure 3: Iso-frequency contours at 2.409GHz for the volumetric NRI-TL metamaterial describing propagation in the layer planes (solid blue curve – obtained through full-wave FEM simulations) and for propagation in free space (red squares). Their near-perfect coincidence suggests that the volumetric NRI-TL metamaterial exhibits a nearly isotropic refractive index of –1 at this frequency.

An iso-frequency surface at 2.409GHz (a difference of less than 0.4% from the design frequency, and, once again, well within the tolerance due to discretization of the FEM mesh) is depicted in Fig. 3 and is shown superposed on the iso-frequency surface corresponding to propagation in free space (the 'light cone'). The coincidence of the two curves suggests that the volumetric NRI-TL metamaterial possesses isotropic $\mu_{\text{eff}}(\omega)=-\mu_0$ and $\varepsilon_{\text{eff}}(\omega)=-\varepsilon_0$ at this frequency for all directions of propagation within the plane, as required by a true Veselago-Pendry superlens.

IV. EXPERIMENT

The fabricated multilayer NRI-TL metamaterial Veselago-Pendry lens is shown in Fig. 4A. The inset depicts the loading on a single layer, from which the reader may make out the CPS TL, series capacitors (oriented at ±45° with respect to the CPS TL axes) and shunt inductors (oriented at 0° or 90° with respect to the CPS TL axes).

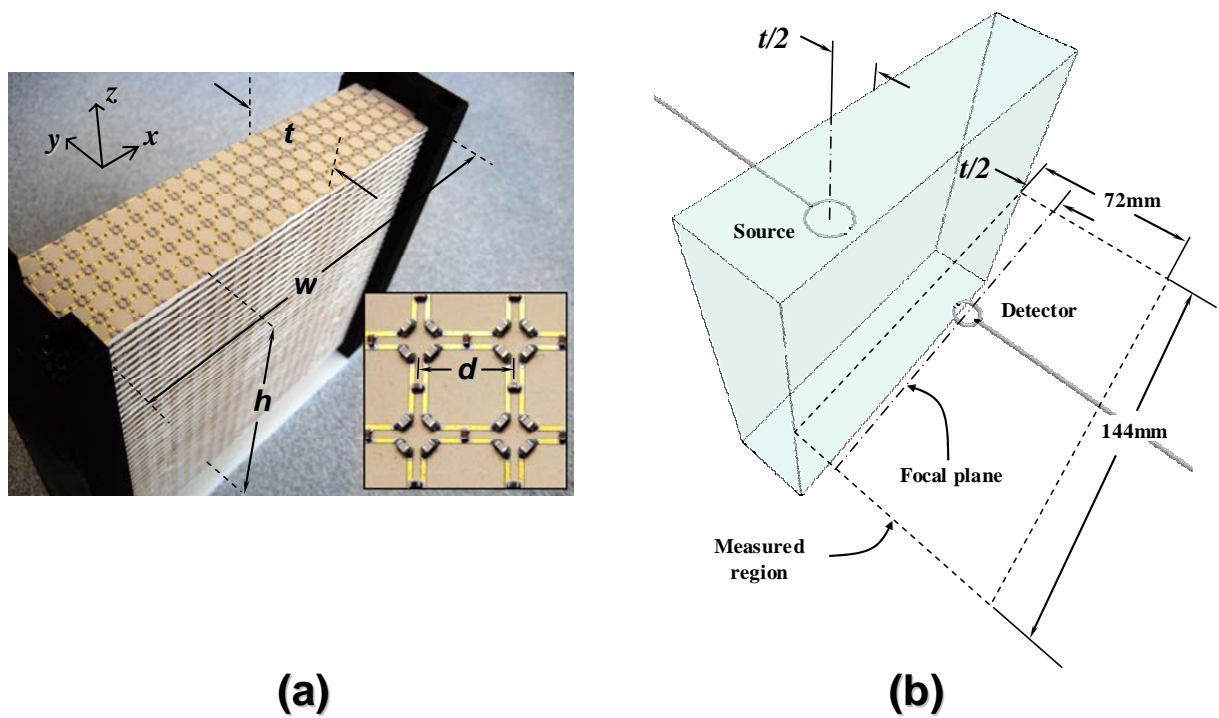

(a)   (b)

Figure 4: a) Photograph of fabricated NRI-TL lens with inset showing lumped loading of the host coplanar-strip TL structure using discrete surface-mount inductors and capacitors. ($w \times h \times t$ = 149.9mm×149.5mm×35.7mm, $d$ = 7.14mm); b) Measurement arrangement – the marquee indicates the region in which the field data presented in the following figures are measured.

The lens consists of 43 layers, each containing a 21×5 array of NRI-TL unit cells. Since each unit cell measures $d$=7.14mm square, the lens measures $w$=149.9mm wide by $t$=35.7mm thick. The layers were held rigidly in place using a plastic frame designed to maintain the layer period of 3.476mm, resulting in a total lens height of $h$=149.5mm. These dimensions are such that a source placed a distance of $t/2$ from the front face of the lens encounters a numerical aperture of 0.97, which collects most of its propagating spectrum (the restoration of the evanescent spectrum is a function of the lens design and the proximity of the source and lens); thus, the physical size of the NRI-TL does not impose severe restrictions on its imaging ability. The transverse dimensions are approximately $1.2\lambda_0$, which provides a sufficient illumination area and minimizes diffraction around the edges.

Following chemical etching of the substrate material and contour routing of the layers, the nearly 26000 surface-mount components were precisely placed using a Siemens SIPLACE assembly and placement system. The boards were then cut manually and inserted into a plastic frame designed to maintain the layer spacing. The measurement apparatus consisted of a source loop antenna (21-mm ≅ $\lambda_0/6$ diameter) and detector loop antenna (7-mm ≅ $\lambda_0/18$ diameter) connected to the terminals of an Agilent E8364B Performance Network Analyzer (PNA). The antennas were constructed from a semi-rigid 50-Ω microwave coaxial cable (1.19-mm outer diameter), and employed a 'shielded' topology that provides magnetic-dipole-type fields while simultaneously minimizing unwanted radiation from unbalanced currents on the coaxial feeding structure [24]. It has been shown that the fields in the volumetric metamaterial remain strongly confined to the layer planes, and so, provided that the fields at the output are measured in the same horizontal plane, the single loop antenna appears to excite the affected layers like an infinite line source [20]. The source (illuminating) antenna was placed at a distance of $t/2$=17.85mm from the front face of the lens, and the detector loop antenna was affixed to a computer-controlled *xyz*-translator and scanned behind the lens in the plane of excitation for field magnitude and phase distributions suggestive of focusing and evanescent decay. The use of a larger loop for the source produces strong fields that are easier to detect, and the use of a small loop for the receiver allows the detection of these fields without disturbing them. The measurement apparatus was covered in a microwave absorber and transmission measurements between the two antennas were taken in intervals of 4mm ($\lambda_0/31$). The data were averaged 30 times to minimize noise in the measurement. Figure 4B shows the measurement arrangement; the marquee identifies the region in which the fields are sampled and represents the region in which the data of Figs. 5–6 are taken.

Figures 5A and 5B present the raw measured field magnitude and phase data over the measurement region at the operating frequency of 2.40GHz for two cases: Fig. 5A shows the results of a control experiment in which the fields in free space are measured before the lens is inserted; Fig. 5B presents measurements over the same spatial region with the lens in place.

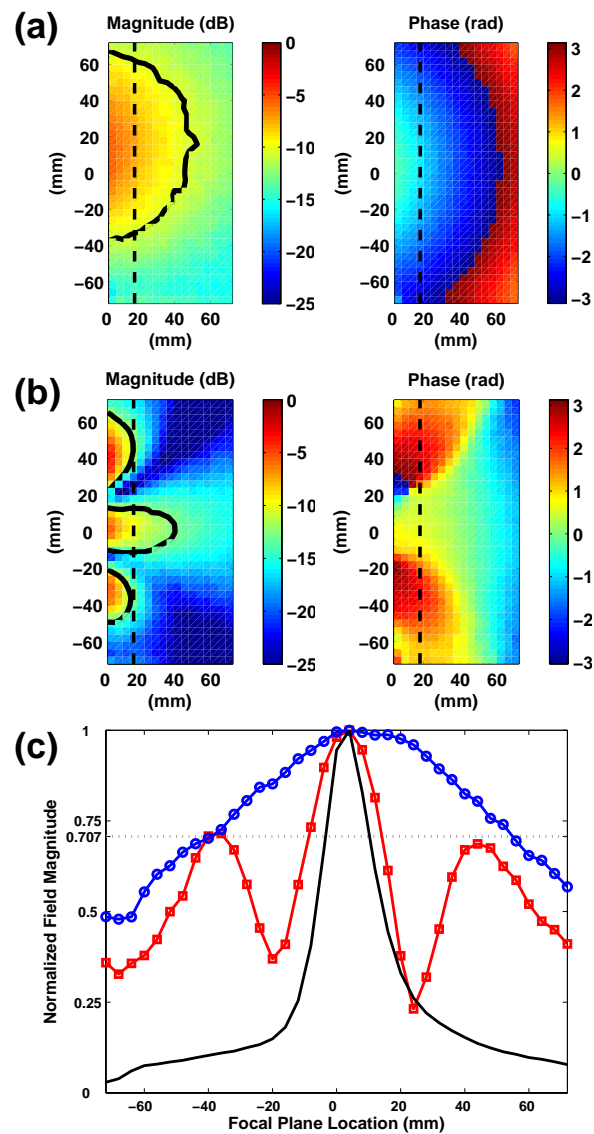

Figure 5: Raw measured magnitude and phase data for excitation with a single-loop source when a) the lens is absent and b) the lens is present. The black curves trace the half-power contours referenced to the maximum field magnitude at the focal plane (dashed line); c) A comparison of the normalized magnitude profiles (linear scale) at the focal plane when the lens is absent (blue circles) and when the lens is present (red squares), along with the fields at a distance of $t/2$ from the source when the lens is absent (solid black curve). The dotted horizontal line indicates the half-power levels and shows that the NRI-TL superlens is able to produce an image of the source with a minimum peak-to-null width of $0.16\lambda_0$.

The black curves indicate half-power contours referenced to the maximum field magnitudes at the expected focal plane, indicated in each case by a dashed line. Since the focal plane lies at a distance of $0.57\lambda_0$ from the source plane, the evanescent-wave components have all but disappeared in the field magnitude distribution of Fig. 5A, and the half-power contour is approximately $0.80\lambda_0$ in width. The nature of the phase fronts in Fig. 5A also suggests that we detect only propagating fields emanating away from a source located at their phase centre. However, the situation is dramatically different when the lens is inserted: the field magnitudes of Fig. 5B indicate the formation of a tightly confined focal region whose transverse half-power width at the focal plane is less than $0.18\lambda_0$ (minimum peak-to-null width of $0.16\lambda_0$), over four times narrower than without the lens and over 3.3 times narrower than that predicted for diffraction-limited images ($0.61\lambda_0$ in free space). The normalized magnitude profiles at the focal plane, as well as the magnitude profile a distance of $t/2$ from the source are compared in Fig. 5C. The evanescent nature of the fields is suggested by the expected decay in the field magnitude distribution of Fig. 5B. It is also suggested by the phase data of Fig. 5B, which remain nearly constant in the focal region and assume a propagating characteristic further from the focal plane, where the strong evanescent fields have decayed and only the propagating fields remain.

To ensure that the resolution ability of the NRI-TL superlens enables the discrimination of two closely spaced sources, the single source antenna was substituted with two identical shielded-loop antennas (each with a 21-mm diameter) fed coherently using a passive microwave power splitter. Due to mutual coupling, the use of practical sources at close range limits the available resolution; in fact, the minimum separation of the sources required to resolve them at their half-power levels even at the front face of the lens, where the decaying evanescent spectrum is collected, was experimentally determined to be between $\lambda_0/3$ and $\lambda_0/4$; the former separation distance was chosen to simplify the analysis of the findings. Figures 6A and 6B present the raw measured field data in the absence and presence, respectively, of the NRI-TL superlens.

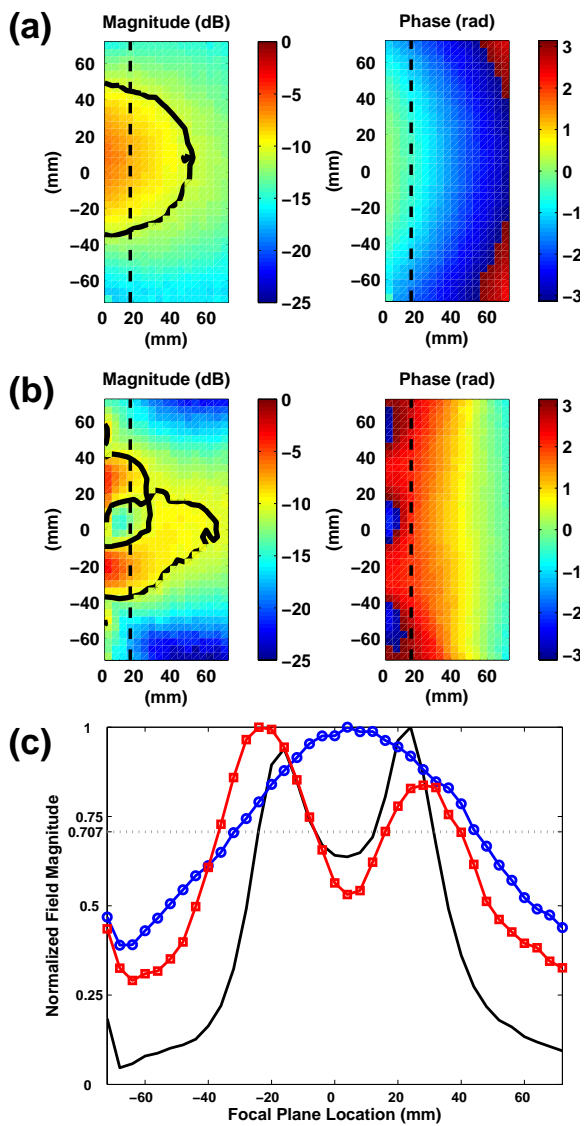

Figure 6: Raw measured magnitude and phase data for excitation with two loop sources separated by $\lambda_0/3$ when A) the lens is absent and B) the lens is present. The black curves trace the half-power contours referenced to the maximum field magnitude at the focal plane (dashed line); C) A comparison of the normalized magnitude profiles (linear scale) at the focal plane when the lens is absent (blue circles) and when the lens is present (red squares), along with the fields at a distance of $t/2$ from the source when the lens is absent (solid black curve). The dotted horizontal line indicates the half-power levels and shows that the NRI-TL superlens is easily able to differentiate the sources.

Once again, the black curves are half-power contours normalized to the maximum field amplitudes at the focal plane (dashed line). Figure 6C presents the normalized magnitude profiles of the two sources at the front face of the lens, at the image plane without the lens in place, and at the image plane with the lens in place. It is evident from these data that the NRI-TL superlens is able to recover the fine distinguishing features that are lost when the lens is absent, and nearly reproduces the available resolution of $\lambda_0/3$. This result represents a resolution ability nearly twice as good as that offered by conventional lenses constrained by the diffraction limit, and further testing of more closely spaced sources promises to reveal an even better resolution ability. The evanescent decay of the images from the exit face of the superlens is also evident. It should be noted that the difference in the levels of the two recovered images can be attributed to slight differences in the construction of the small antennas, and slight horizontal misalignment between the two sources and also between the source and detector as the fields are scanned.

From its very early stages, this work was motivated by the desire to see metamaterial superlenses applied to the imaging of small scatterers at practical focal distances, as in close-range non-invasive tumour detection or land-mine detection. Although the experimental results presented so far pertain to luminous sources, the detection of non-luminous objects by way of backscattered fields may also be facilitated by a metamaterial superlens. For example, it has been proposed that a single luminous source at the front of the lens can be used both to illuminate a scatterer behind the lens and detect it by way of its backscattered secondary fields [25]. Alternatively, it is possible to illuminate the front side of the lens using a normally incident plane wave. Since the Veselago-Pendry lens does not possess a unique principal axis, such a plane wave passes directly through the lens and impinges upon the scatterer, whose backscattered fields are then refocused at the front side of the lens where they may be detected. Both cases are depicted in Fig. 7, where the red solid arrows indicate the directions of the illuminating fields and the blue open arrows indicate those of the backscattered fields. Since the antennas employed would necessarily be small, matching techniques or time-gating could be employed to isolate the desired backscattered signals.

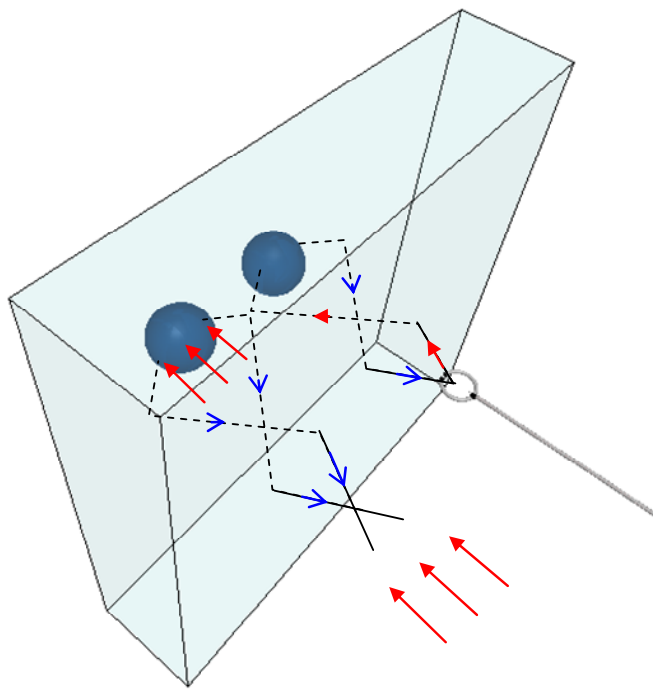

Figure 7: Detection of non-luminous objects behind the lens via backscattered fields. The objects may be illuminated either by the detecting antenna or by a normally incident plane wave, which passes directly through the lens without focusing. The red arrows indicate the directions of the illuminating fields, and the blue arrows indicate those of the backscattered fields.

## V. Conclusion

We have presented experimental evidence of free-space imaging beyond the diffraction limit at 2.40GHz using a Veselago-Pendry superlens based on a multilayer implementation of the NRI-TL metamaterial. The NRI-TL approach affords precise control over the material parameters of the lens and is much less susceptible to losses than other methods. Moreover, this approach is supported by a good agreement between an equivalent-circuit theory and full-wave simulations. The experimental results reveal focusing of a single source to a minimum peak-to-null beamwidth of less than one-sixth of a wavelength and a resolution of two sources displaced transversely by a distance of one-third of a wavelength, both well below the classical diffraction limit. The multilayer NRI-TL implementation is attractive because its constituent layers may be easily and rapidly fabricated using existing PCB fabrication techniques and facilities. This simplified approach to the realization of Veselago-Pendry superlenses should encourage their application to imaging problems in biomedicine, microelectronics, and defense.

ACKNOWLEDGMENT

The authors would like to thank Rogers Corporation, Saturn Electronics, and Jamie McIntyre and Malcolm Forge of George Brown College for their help in fabricating the components of the NRI-TL metamaterial superlens. Financial support from the Natural Sciences and Engineering Research Council of Canada (NSERC) is also gratefully acknowledged.

REFERENCES

[1] V. G. Veselago, "The electrodynamics of substances with simultaneously negative values of $\varepsilon$ and $\mu$," *Sov. Phys. Usp*, vol. 10, no. 4, pp. 509-514, Jan.-Feb.1968.

[2] J. B. Pendry, "Negative refraction makes a perfect lens," *Phys. Rev. Lett.*, vol. 85, no. 18, pp.3966–3969, October 2000.

[3] A. Grbic and G. V. Eleftheriades, "Overcoming the diffraction limit with a planar left-handed transmission-line lens," *Phys. Rev. Lett.*, vol. 92, p. 117403, March 2004.

[4] P. Alitalo, S. Maslovski, and S. Tretyakov, "Three-dimensional isotropic perfect lens based on *LC*-loaded transmission lines," *J. Appl. Phys.*, vol. 99, p. 064912, March 2006.

[5] S. M. Rudolph, A. Grbic, "Volumetric negative-refractive-index medium exhibiting broadband negative permeability," *J. App. Phys.*, vol. 102, p. 013904, Jul. 2007.

[6] M. Stickel, F. Elek, J. Zhu, G. V. Eleftheriades, "Volumetric negative-refractive-index metamaterials based upon the shunt-node transmission-line configuration," *J. App. Phys.*, vol. 102, p. 094903, Jul. 2007.

[7] N. Fang, H. Lee, C. Sun, X. Zhang, "Sub-diffraction-limited optical imaging with a silver superlens," *Science* vol. 308, no. 5721, pp. 534–537, Apr. 2005.

[8] F. Mesa, M. J. Freire, R. Marquès, J. D. Baena, "Three-dimensional superresolution in metamaterial slab lenses: Experiment and theory," *Phys. Rev. B*, vol. 72, p. 235117, Dec. 2005.

[9] M. C. K. Wiltshire, "Radio-frequency (RF) metamaterials," *Phys. Status Solidi B*, vol. 244, no. 4, pp. 1227–1236, Mar. 2005.

[10] A. Salandrino, N. Engheta, "Far-field subdiffraction optical microscopy using metamaterial crystals: Theory and simulations," *Phys. Rev. B*, vol. 74, p. 075103, Aug. 2006.

[11] Z. Jacob, L. V. Alekseyev, E. Narimanov, "Optical Hyperlens: Far-field imaging beyond the diffraction limit," *Opt. Express*, vol. 14, no. 18, 8247, Sep. 2006.

[12] Z. Liu, H. Lee, Y. Xiong, C. Sun, X. Zhang, "Far-field optical hyperlens magnifying sub-diffraction-limited objects," *Science*, vol. 315, no. 5819, p. 1686, Mar. 2007.

[13] K. Aydin, and E. Ozbay, "Focusing of electromagnetic waves by a left-handed metamaterial flat lens," *Optics Express*, vol. 13, no. 22, p. 8759, Oct. 2005.

[14] K. Aydin, I. Bulu, E. Ozbay, "Subwavelength resolution with a negative-index metamaterial superlens," *App. Phys. Lett.*, vol. 90, p. 254102, Jun. 2007.

[15] D. R. Smith, D. Schurig, M. Rosenbluth, S. Schultz, S. A. Ramakrishna, J. B. Pendry, "Limitations on subdiffraction imaging with a negative refractive index slab," *App. Phys. Lett.*, vol. 82, pp. 1506–1508, Mar. 2003.

[16] R. Marquès, M. J. Freire, and J. D. Baena, "Theory of three-dimensional subdiffraction imaging," *App. Phys. Lett.*, vol. 89, p. 211113, Nov. 2006.

[17] G. V. Eleftheriades, A. K. Iyer, P. C. Kremer, "Planar negative refractive index media using periodically L-C loaded transmission lines," *IEEE Trans. on Microwave Theory and Tech.*, vol. 50, no. 12, pp. 2702–2712, Dec. 2002.

[18] A. K. Iyer and G. V. Eleftheriades, "A volumetric layered transmission-line metamaterial exhibiting a negative refractive index," *J. Opt. Soc. Am. B.* (*JOSA-B*) *Focus Issue on Metamaterials*, vol. 23, pp. 553–570, Mar. 2006.

[19] A. K. Iyer and G. V. Eleftheriades, "Characterization of a multilayered negative-refractive-index transmission-line (NRI-TL) metamaterial," *IEEE MTT-S Int. Microwave Symp. Dig*. (San Francisco, CA, USA), pp. 428-431, June 11–16, 2006.

[20] A.K. Iyer, G.V. Eleftheriades, "A multilayer negative-refractive-index transmission-line (NRI-TL) metamaterial free-space lens at X-band," *IEEE Trans. Antennas and Propagat.*, vol 55, no. 10, pp. 2746–2753, Oct. 2007.

[21] G.V. Eleftheriades, "Analysis of bandwidth and loss in negative-refractive-index transmission-line (NRI–TL) media using coupled resonators," *IEEE Microwave and Wireless Comp. Lett.*, vol 17, no. 6, pp. 412–414, Jun. 2007.

[22] Ansoft Corporation, http://www.ansoft.com/products/hf/hfss/

[23] D. R. Smith, S. Schultz, P. Markŏs, C. M. Soukoulis, "Determination of effective permittivity and permeability of metamaterials from reflection and transmission coefficients," *Phys. Rev. B*, vol. 65, no. 19, p. 195104, May 2002.

[24] "The Loop Antenna for Transmission and Reflection" by R. W. King in Antenna Theory: Part I, Eds. R. E. Collin and F. J. Zucker, Ch. 11, 478–480, *McGraw-Hill* (Toronto: 1969)

[25] G. Wang, J. R. Fang, X. T. Dong, "Refocusing of backscattered microwaves in target detection by using LHM flat lens," Opt. Express, vol. 15, no.6, 3312–3317 , Mar. 2007.